\documentclass[article]{aa}

\usepackage{graphics}
\usepackage{epsfig}
\usepackage{latexsym}
\begin{document}
\hyphenation{brems-strah-lung}
\title{The 2008 outburst of IGR~J17473--2721: \\
evidence for a disk corona?\thanks{Based on observations with INTEGRAL, an ESA project with instruments and
science data centre funded by ESA member states (especially the PI
countries: Denmark, France, Germany, Italy, Switzerland, Spain), Poland
and with the participation of Russia and the USA.}}

\author{Yu-Peng Chen\inst{1}, Shu Zhang\inst{1},  Diego F. Torres\inst{2},  Shuang-Nan Zhang\inst{1}, \\
Jian Li\inst{1}, Peter Kretschmar\inst{3}, Jian-Min Wang\inst{1,4}}

\institute{Laboratory for Particle Astrophysics, Institute of High
Energy Physics, Beijing 100049, China
\and
ICREA \& Institut de Ci\`encies de l'Espai (IEEC-CSIC), Campus UAB,
Facultat de Ci\`encies, Torre C5-parell, 2a planta, 08193 Barcelona,
Spain
\and
ISOC, ESA/ESAC, Urb. Villafranca del Castillo, PO Box 50727, 28080 Madrid, Spain
\and
Theoretical Physics Center for Science Facilities (TPCSF), CAS
          }

\offprints{Yu-Peng Chen}
\mail{chenyp@mail.ihep.ac.cn}

\date{ }

\date{Received  / Accepted }

\titlerunning{The 2008 outburst of IGR~J17473--2721}
\authorrunning{Chen et al.}

  \abstract
  {The 2008 outburst  of the atoll source IGR~J17473--2721 was  observed by \emph{INTEGRAL},\emph{RXTE} and \emph{Swift}. Tens of type-I X-ray bursts were found in this outburst. }
{Joint observations  provide sufficient data to look into the  behavior of
IGR~J17473--2721 at the rising part of the outburst. The relation between the duration of the bursts and
the accretion rate and the nature of the corona producing the observed power-law component can therefore be
studied in detail. }
{We analyze  observational data of IGR J17473--2721, focusing on the
spectral evolution during the state transition from quiescent to low hard state (LHS),   and on the flux dependence of
the  type-I X-ray bursts along the outburst.
 }
{We find that  the joint energy spectra can be well fitted with a model composed of a blackbody and a cutoff power-law, with a cutoff energy decreasing from $ \sim$ 150 keV to $\sim$ 40 keV as  the source leaves the quiescent state toward the low hard state.
This  fits into a scenario in which the corona is cooled by the soft X-rays along the outburst evolution, as observed in several other atoll sources.
By using the flux
measured in the 1.5--30 keV band of the type-I bursts during the outburst,  we find that the linear relationship between the burst duration and the flux still
holds for those bursts that occur at the decaying part of the low hard state, but with a different slope  than the overall one that was  estimated with the bursts happening in the whole extent of,  and for the rest of the low
hard state. The significance of such a  dichotomy in the type-I X-ray bursts is $\sim$ 3 $\sigma$ under an F-test.  Similar results are hinted at as well  with the broader energy-band that was adopted recently. This  dichotomy  may be understood in a scenario where part of the accreting material   forms a corona on the way
of falling onto the surface of the neutron star during the decaying part of the low hard state. Based on the accretion rates of the preceding LHS, estimated from  type-I X-ray bursts and  from  persistent emission, at least for IGR J17473-2721, most of the accretion material may fall on the neutron star (NS)
surface in the LHS. Considering the burst behavior  in the context of the outburst indicates a corona formed  on top of the disk rather than on the NS
surface.
}
{} 

   \maketitle

\section{Introduction}
Low-mass X-ray binaries (LMXB) containing neutron stars (NSs) are classified as atoll or Z sources according to their different   spectral and timing evolution during outburst (Hasinger and van der Klis 1989).
Atoll sources have a typical luminosity $L_{X}$ $<$ 10\%
Eddington, and mostly evolve from  the island  to the  banana state, analogous to that of the black hole (BH) counterparts, which evolve from the LHS to the high/soft states (HSS).  Z sources are usually brighter,  with softer spectra and weaker variability, in analogy to the HSS in BH binaries.

In general, the spectra of LMXB (both BH and NS binaries) are described by two-component models, including a soft/thermal (e.g., a blackbody) and a hard/Comptonized component  (e.g., power-law shape with a cutoff at  tens to hundreds of keV).
It is  believed that the hard component evolves in the inverse Compton scattering of soft/thermal photons off the hot electrons in the corona. With respect to the corona itself,  two main uncertainties remain for  both black hole and NS systems:    how are the electrons in the corona  energized up to $\sim$ 100 keV, and what is the corona geometry and location (Reynolds \&  Nowak 2003). Regarding the latter,  the Eastern (Mitsuda et al. 1989) and the Western models (White et al. 1988)  have been proposed for NS binaries; whereas models with the so-called  ``sandwich" geometry, ``sphere+disk geometries", and ``patchy corona"  were proposed as well for BH binaries (see Figure 6 of  Reynolds \&  Nowak 2003). The Eastern model considers that the thermal emission arises from the disk and that the Comptonization arises from hot plasma around the neutron star. Instead, the Western model considers that the thermal emission arises from the boundary layer and that  the Comptonization arises from the disk.

For the Comptonized components  of both BH and NS binaries in LHS, the high-energy cutoff is present at tens to hundreds of keV. This can be an indicator of the  electron  Maxwellian energy distribution in the  corona and  approximately twice  the electron temperature (Hua \& Titarchuk 1995). Studying the evolution of the cutoff energy during outbursts is essential for understanding the accretion processes.
For the LHS of BH binaries, the power-law is cut at $\sim$ 100 keV, while no cutoff was detected up to 10 MeV
in the HSS of seven transient BH candidates: GRO~J0422+32, GX~339-4, GRS~1716-249, GRS~1009-45, 4U~ 1543-47,
GRO~J1655-40, and GRS~1915+105 (Grove et al. 1998). Recently, for the well-studied BH binary GX~339-4, the
cutoff energy monotonically decreased, starting from $\sim$200 keV and reaching $\sim$60 keV, while the flux of
the whole X-ray band increased during the LHS before the HSS (Motta et al. 2009). Interestingly, the cutoff
energy is still detected (at $\sim$ 100 keV) in high-intermediate and soft-intermediate states and suddenly
disappears in the HSS state.
For the NS binaries, no cutoffs were observed in the bright Z sources (see the review by Done et al. 2007).
For some of the atoll sources observed, the cutoff energy is seen at more than 20 keV in LHS (Barret \& Vedrenne 1994). For instance,
Chen et al. (2006) found that the cutoff energy changed from $\sim$ 3 keV to $>$ 30 keV during the spectral transition from the HSS to the
LHS of 4U 1608-522, proposing that the existence of a
solid surface or the influence of a strong magnetic field may have a great impact on this difference with BH binaries.

However, the changes in the cutoff energy of NS binaries while the source transitions from quiescence  to the LHS are
much less addressed, partially because the fact that the outbursts are usually caught when the source leaves the
quiescent state where  the emission dominated by hard X-rays is too weak.  The joint observations of the initial
rising phase of the outburst from IGR J17473--2721 by \emph{INTEGRAL} and \emph{RXTE} provide a nice opportunity
to conduct this study. The atoll source IGR~J17473--2721 has shown a variety of spectral states/transitions in
the 2008 outburst. At the rising phase of this outburst, the source showed hard X-ray flares and stayed in the
low/hard state for two months prior to transition to the high/soft state, forming the so-called hysteresis.
IGR~J17473--2721 is thus the third NS binary with hysteresis after 4U~1908+005 (Aql X-1) and 4U~1608-522 (e.g.
Gladstone et al. 2007).
The outburst with a low/hard state prior to a high/soft state resemble Aql~X-1 and XTE J1550-564. The long-lived
preceding LHS makes IGR J17473--2721 to resemble the behavior of outbursts seen in black hole X-ray binaries
like GX 339-4.
  Zhang et al. (2009) presented details of RXTE observations on the
outburst, but
  data for the initial part were not yet available.

Type-I X-ray bursts manifest themselves as a sudden increase (typically  by a factor of 10 or greater) in the X-ray
luminosity of neutron star systems (NSs), with $\sim$90 X-ray bursters detected in our Galaxy (Grindlay et al.
1976, Belian et al. 1976). Type-I bursts are caused by unstable burning of accreted H/He on the surface of
neutron stars in low-mass X-ray binary (LMXB) systems, in contrast to type-II bursts, that are thought to be
caused by accretion instability (for reviews, see Lewin et al. 1993, Cumming 2004,  Strohmayer \& Bildsten 2006,
and Galloway et al. 2008).
The properties of type-I X-ray bursts depend on the composition of the burning material, how stable the
burning of that fuel is, and if there is any left-over fuel from previous bursts.
%
Several other factors can also significantly change the burst properties: individual element reaction rates, fuel  mixing in
the burning layer, inhomogeneous distribution of fuel, and other aspects.

Tens of type-I bursts were detected in IGR~J17473--2721  during  outburst by \emph{RXTE}, \emph{INTEGRAL},
Swift, and SuperAGILE (Altamirano et al. 2008, Galloway et al. 2008, Chen et al. 2010, Chenevez  et al. 2010).
Chen et al. (2010), who found 16 bursts, estimated that the distance is likely to be  6.4$\pm$0.9 kpc based on
the peak flux of three bursts that show photospheric radius expansion. The duration of the 16 type-I bursts
occurred in the 2008 outburst is found to correlate with the Eddington ratio and with two parallel evolution
groups (Chen et al. 2010). Analysis of more complete RTXE data by Chenevez  et al. (2010)  revealed  that 57
bursts (42 bursts among  those detected by RXTE) were born in the 2008 outburst of IGR J17473--2721, with an
overall correlation between the burst duration and the flux.   However, Chenevez  et al. (2010)  adopted an
expanded energy band (0.1--200 keV) to calculate the bolometric flux of the persistent emission, far beyond
\emph{RXTE}'s best  coverage (1.5--30 keV), which may result in flux uncertainties as large as  $\sim$40\%.
This stimulates us to re-investigate the duration/flux relation with this burst sample. Finally, \emph{INTEGRAL}
data on the initial rising part of the outburst were never reported before, which could provide clues on the
formation of the outburst from quiescence and help investigating the initial cooling process of the corona. We
report on this here as well.

\section{Observations and data analysis}

\begin{figure}[t]
\centering
\includegraphics[angle=0, scale=0.4]{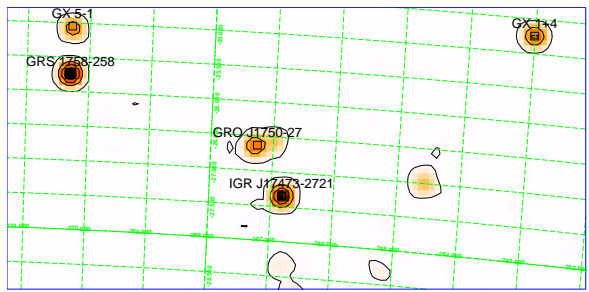}
     \caption{
    Significance map  of IGR~J17473--2721  during the outburst in 2008, as observed by INTEGRAL. The contour start at a significance level of 5$\sigma$ with steps of 50$\sigma$.
}
         \label{image}
\end{figure}

\begin{figure*}
\centering
  \includegraphics[angle=0, scale=0.7]{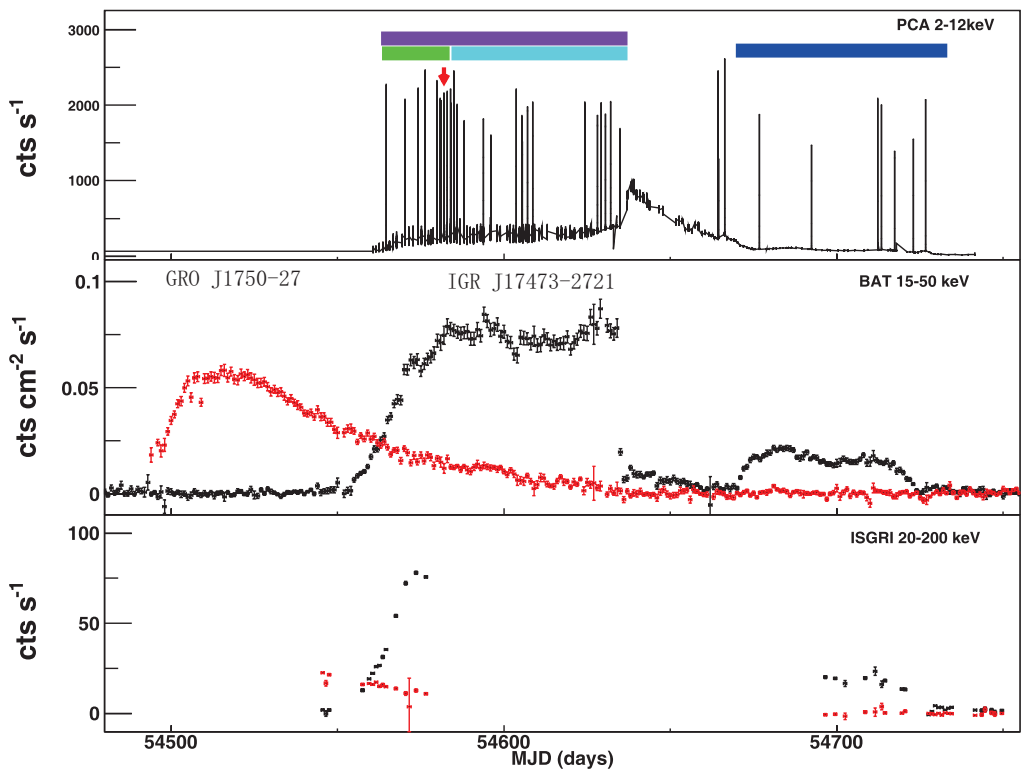}
     \caption{
     Upper panel: PCA (2-12 keV) lightcurve of IGR~J17473--2721 with a time bin of 16 seconds, in which the spikes are type-I X-ray bursts.
The color bar above the PCA lightcurve indicates the burst groups for the subsequent analysis of the
relationship between the burst duration and the corresponding persistent flux. Middle panel: the BAT (15-50 keV)
lightcurve  of IGR J17473--2721 (black) and GRO J1750$-$27 (red) during the outburst in 2008. Lower panel: the
ISGRI (20-200 keV) lightcurves of both sources monitored by \emph{INTEGRAL}. In both of the lower panels the
bin size is one day. }
         \label{image2}
\end{figure*}

The available \emph{INTEGRAL} observations, when IGR~J17473--2721 had an offset angle less than 14$^\circ$ during
the outburst in 2008, comprised about 463 SCWs, adding up to an exposure time of $\sim$ 1.3 Ms (covering
revolutions 0663--0723 --one revolution covers $\sim$three days-- between MJD 54545.1 and MJD 54750.0). However, data
only scatter over the rising and decay phases in this outburst, and only the former is used here because of the
low significance of detection in the decay phase. The data reduction was performed using the OSA version 9.0.
ISGRI and JEMX data are used to extract the spectra and lightcurves. All  sources within the FoV that are brighter
than, or comparable to, IGR~J17473--2721 were taken into account in extracting the source spectrum.

Public data from \emph{RXTE} (Gruber et al. 1996) on IGR~J17473--2721 cover the entire 2008 outburst
 between April 2008 and October 2008, and include 182 \emph{RXTE}/PCA pointed observations,
with the identifier (OBSID) of proposal number (PN) 93064, 93093, and 93442 in the
 High Energy Astrophysics Science Archive Research Center (HEASARC).
These observations add up  to $\sim$ 475 ks of exposure time on the source, and scatter over the entire outburst.
The analysis of the PCA data was performed by using
 HEAsoft v. 6.6.  We filtered the data using the standard \emph{RXTE}/PCA criteria.
 Only the PCU2 (in the 0-4 numbering scheme) was used for the analysis, because  this PCU was 100\% working on during the
 entire observations.
The background file used in the analysis of PCA data is the most recent one for
bright-sources  found at the HEASARC website\footnote{
pca$\_$bkgd$\_$cmbrightvle$\_$eMv20051128.mdl};
the detector breakdowns have been removed.
Data from cluster 1 of the HEXTE system were used to produce the  higher energy spectra.

An additional 1\% systematic error is added to the
spectra because of calibration uncertainties, if not otherwise specified.  The  X-ray
spectra are fitted with XSPEC v12.5.0  and the model parameters are
estimated  with a 90$\%$ confidence level.
In fitting the joint  \emph{RXTE} \& \emph{INTEGRAL} data points, the JEMX, HEXTE, and ISGRI spectra were normalized to the
PCA one.

There is only a single OBSID 00031182001 from \emph{Swift}  on IGR~J17473--2721 during the 2008 outburst, on 2008 March 3 (MJD 54556).
During the observation, the XRT data  were performed  with  Windows Timing (WT) mode and Photon Counting (PC) mode.
We extracted  PC mode events using the tool XSELECT V2.4a as a part of HEASOFT v6.9 package, and
pile-up effects were corrected.


\section{Results}

\subsection{Outburst}

We find that IGR~J17473--2721 has a detection significance of $\sim$182$\sigma$ in the 20--200 keV band during
the period between MJD 54545.1 and 54750.0 (see the INTEGRAL map in Fig. \ref{image}). The nearby source
GRO~J1750-27, that is  48.92 arcmin away from IGR J17473--2721, was out of the \emph{RXTE}/PCA field of view
during the observations.
The field of view of RXTE/PCA is $\sim$1 degree, RXTE/PCA can't distinguish two sources separated by $\sim$ 1 degree if one is located at the center of the field. RXTE/PCA can avoid contamination from GRO~J1750-27 by pointing   $\sim$0.2 degree away from IGR~J17473-2721.
Fig. \ref{image2} shows the BAT \footnote{See the Swift/BAT transient monitor results provided by the Swift Team at http://swift.gsfc.nasa.gov/docs/swift/results/transients} and ISGRI lightcurves of both sources in bins of one day.

\begin{figure}
\centering
 \includegraphics[angle=0, scale=0.4]{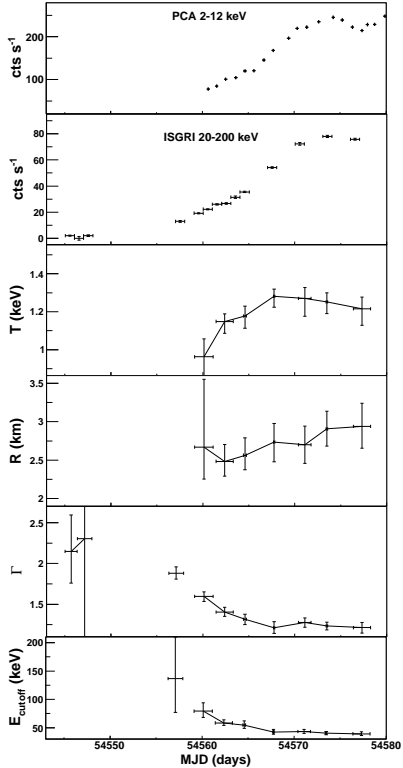}
      \caption{
       Top two panels: \emph{RXTE}/PCA lightcurves of IGR J17473--2721 (2--12 keV, time bin is fixed to one obsid) and \emph{INTEGRAL}/ISGRI, 20-200 keV, time bin is fixed to one day) during the rising phase of its outburst in 2008.
The temperature ($T$) and radius ($R$) of the blackbody, the photon index ($\Gamma$), and the cutoff energy resulting from fit results with \emph{Swift}/XRT (0.5-10 keV),  \emph{RXTE}/PCA (3-30 keV), \emph{RXTE}/HEXTE (30-100 keV), \emph{INTEGRAL}/JEMX (5-25 keV), and \emph{INTEGRAL}/ISGRI (20-200 keV) are also given in the remaining four panels
(the time bin of the bottom panels is one revolution, roughly happening every three days).}
         \label{lc_flux}
\end{figure}

We extracted the \emph{RXTE}/PCA lightcurves in 2--12 keV with PCA standard 2 model and binned it into
observation identifiers (OBSID), with the background removed. This is shown in the top panel of Fig.
\ref{lc_flux}, where for comparison the \emph{INTEGRAL}/ISGRI lightcurves in 20--200 keV are also shown. From
Fig. \ref{lc_flux} it is possible to see that IGR~J17473--2721  left from quiescent state and started to be
active at around MJD 54555, with the flux increasing in both energy bands. Roughly 20 days later, at  MJD
$\sim$54575, the source   stayed  at the LHS and the corresponding results are reported in Zhang et al. (2009).

To analyze the spectral evolution of the initial rising phase of the outburst from IGR~J17473--2721, the ISGRI data
were  divided into ten groups based on the revolution number: rev. 663, 664, 667--674. For each group we fitted the
spectra in the 20--200 keV band with a power-law, a cutoff power-law, or cutoff power-law, both with 1\% systematic error included.
No \emph{RXTE} or \emph{Swift} observations are available during the first two revolutions (663 and 664), and
the source is too weak to have a meaningful JEMX spectrum.
There is one \emph{Swift} observation quasi-simultaneous to the \emph{INTEGRAL} revolution 667, therefore  the joint
\emph{Swift}/XRT, \emph{INTEGRAL}/JEMX and \emph{INTEGRAL}/ISGRI was derived.
IGR~J17473-2721 is outside the  FoV of JEMX during rev. 668,  the joint spectrum was obtained from
\emph{INTEGRAL}/ISGRI, \emph{RXTE}/PCA and \emph{RXTE}/HEXTE observations.
For rev. 669--674, we fitted the combined
\emph{INTEGRAL}/JEMX, \emph{INTEGRAL}/ISGRI, \emph{RXTE}/PCA and \emph{RXTE}/HEXTE spectrum.  \emph{RXTE} OBSIDs where bursts occurred
were excluded from the spectral analysis.

We extracted for each data
group the spectra from the combined  XRT data in 0.5-10 keV data, the PCU 2 data in the 3--30 keV band, the JEMX data in the 5-25 keV band,
the HEXTE cluster 1 data in the 30--100 keV band and ISGRI data in the 20--200 keV band. We used the blackbody model (bbodyrad in Xspec) to fit the thermal component --because the   disk emission is hardly visible owing to the relatively steep disk inclination angle (Zhang et al. 2009)-- and a cutoff power law for the hard component(Table \ref{tab_outburst}).
The iron line is fixed at 6.4 keV and the absorption is set to 3.8$\times 10^{22}$ atoms/cm$^{2}$. A constant is introduced to account for the   normalization difference among XRT, PCA, JEMX, HEXTE and ISGRI.

\begin{figure}[ptbptbptb]
\centering
\includegraphics[angle=-90, scale=0.4]{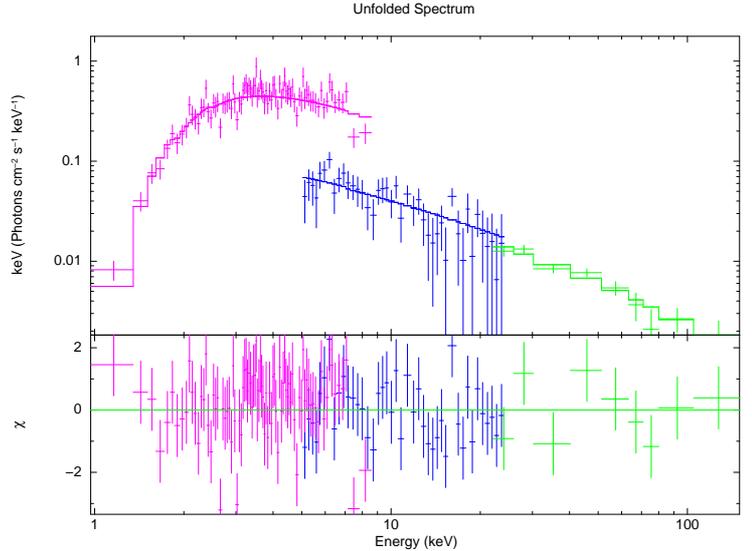}
     \caption{Joint XRT (purple), JEMX (blue), ISGRI (green) spectra of rev. 667 of IGR~J17473--2721 during the outburst in 2008, fitted with model: wabs*cutoff.
}
         \label{spe_667}
\end{figure}

\begin{figure}[ptbptbptb]
\centering
\includegraphics[angle=-90, scale=0.4]{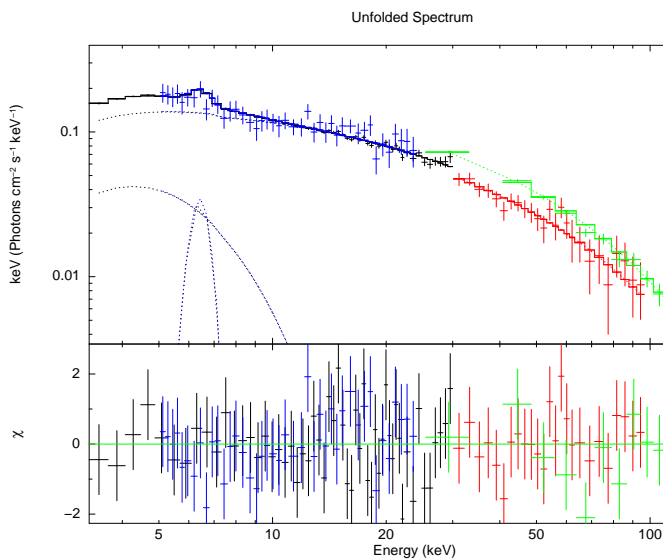}
     \caption{Joint  PCA (black), JEMX (blue), HEXTE (red), ISGRI (green) spectra of rev. 674 (the last revolution of \emph{INTEGRAL} in the rising phase) of IGR~J17473--2721 during the outburst in 2008, fitted with model wabs*(bbodyrad+cutoff+Gauss).
}
         \label{spe}
\end{figure}

We find that at the beginning of the outburst (rev. 663 and 664 between MJD 54545 and 54550), and where only ISGRI  spectra are available,  a power law with a photon index of $\sim$ 2.2 (typical error of 0.2-1.0) can   fit the spectra in the 20--200 keV band well, with reduced $\chi^{2}$ 1.6, 1.2 respectively (d.o.f. is 7, 7).
For rev. 667 (MJD 54556), the joint XRT/ISGRI/JEMX data can be  fitted with an absorbed cutoff power law model, resulting in a  $\chi_{\rm red}^{2}$  (1.11 for 134 d.o.f.) comparable to that derived with a simple power law model (1.15 for 135  d.o.f.) (Fig. \ref{spe_667}, Table \ref{tab_outburst}).
By adding  a  blackbody and a Gaussian line fixed at 6.4 keV  as additional components, the fit can be slightly improved ($\chi_{\rm red}^{2}$/d.o.f. $\sim$1.04/130), but the F-test shows a probability of only 0.022.



From rev. 668 (MJD 54559) onward,
the thermal component can be satisfactorily fitted with a blackbody model, while the fit of the hard component by a simple power law turns worse, with a reduced $\chi^{2}$ larger than 2 (Table \ref{tab_outburst}).
 Adopting the hard component as a
a cutoff power law results in better fits, with  reduced $\chi^{2}$ improving to $\sim$ 1 (e.g., Fig. \ref{spe} shows the spectral fit during rev. 674 as an example). 
During the initial rising phase of the outburst we find that the temperature of the soft emission rises from $\sim$1.0 to $\sim$1.3 keV, and the radius of blackbody slowly increases  from $\sim$2.5 km
to $\sim$2.7 km, the latter values are compatible to those derived from the subsequent low/hard state as reported in Zhang et al. (2009). As the source left  the quiescent state and started to be active, the  cutoff  energy of the hard component decreased from $\sim$136 keV to $\sim$40 keV, with the latter being consistent with that derived in the subsequent low/hard state as reported in Zhang et al. (2009).
The confidence contour plots of $\Gamma$  and cutoff energy are given in  Fig.~\ref{contour} for three INTEGRAL revolutions. While some correlation between the two parameters is evident in the
  elliptical shapes of the contours, the different fit results are nicely
separated,
  supporting the reality of the trend shown in the lower two panels of Fig. \ref{lc_flux}.

\begin{table}[ptptptpt]
\begin{center}
\label{tab_outburst}
\caption{The spectral fit results of the outburst with different models. }
\begin{tabular}{c|ccccccccccccccccc}

\hline \hline
rev                       &powerlaw  & cutoff  & cutoff+bbodyrad+gauss\\\cline{2-4}
                               &\multicolumn{3}{c}{$\chi_{\rm red}^{2}$(d.o.f)}\\\hline
663$^a$                      &1.6(7)      &1.96(6)            &-\\
664$^a$                      &1.2(7)      &1.45(6)            &-\\\hline
667$^b$            &1.15(135)   &1.11(134)   &1.04(130)\\\hline
668$^c$             &5.22(97)    &3.91(96)    &1.30(92)\\\hline
669$^d$       &7.33(134)   &3.45(133)   &1.04(129)\\
670$^d$        &4.67(138)   &2.95(137)   &0.67(133)\\
671$^d$        &6.26(138)   &2.39(137)   &0.68(133)\\
672$^d$        &10.1(137)   &4.10(136)   &0.84(132)\\
673$^d$        &12.9(138)   &3.97(137)   &0.71(133)\\
674$^d$        &11.1(135)   &3.22(134)   &0.83(130)\\\hline
\end{tabular}
\end{center}
\begin{list}{}{}
\item[${^{a}}$]{with ISGRI data}
\item[${^{b}}$]{with ISGRI+JEMX+XRT data}
\item[${^{c}}$]{with ISGRI+PCA+HEXTE data }
\item[${^{d}}$]{with ISGRI+JEMX+PCA+HEXTE data}
\end{list}
\end{table}

\subsection{Type-I X-ray bursts}

In Chen et al. (2010) we have analyzed 16 bursts that occurred during the 2008 outburst of IGR J17473--2721 and
found that they could be grouped into two parallel evolution  tracks regarding their burst duration versus their
flux. However, such parallel tracks did not show up in the work by Chenevez  et al. (2010), where 57
bursts were found in the entire outburst, although  the overall trend of a positive relationship still holds.
Instead of taking the 0.1--200 keV band  by Chenevez  et al. (2010) did,
we adopt a band  well covered by RXTE (1.5--30 keV)  to minimize the uncertainties intrinsic to flux extrapolations (see, e.g., Thompson et al. 2008).

\begin{figure*}
\centering
 \includegraphics[angle=0, scale=0.7]{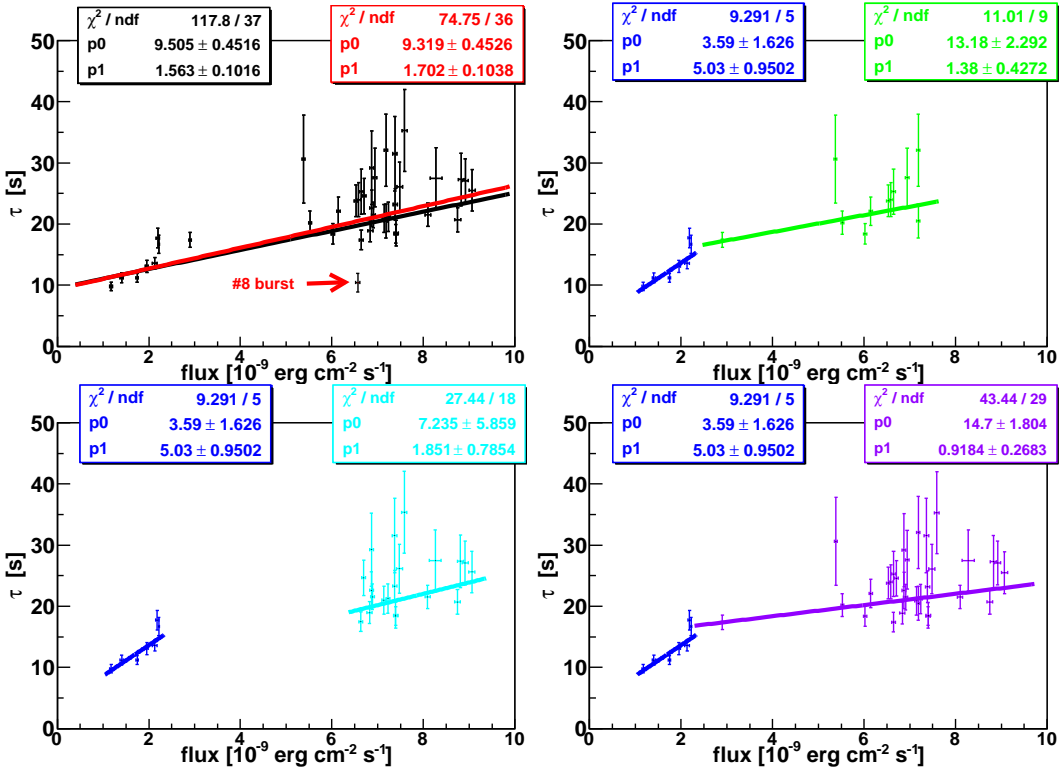}
      \caption{Timescale $\tau$ vs. the persistent flux in the 1.5--30 keV band for the bursts occurring in LHS.
The black
points represent all of the bursts together (upper left panel). For comparison, each burst group is plotted together with the results of a linear fit to this group
marked by
   different colors.
The top-left panel shows the  fit results for all  bursts (with larger reduced $\chi^{2}$, black) and all   bursts except the burst number 8 (in red).
The top-right panel shows the  fit results of the bursts during LHS after the HSS (blue), and the bursts occurring during the rising part of the LHS before the HSS (green).
The bottom-left panel presents the fit results of the bursts happening at the plateau of the LHS before the HSS and
the bottom-right panel those for the bursts  occurring during the LHS before the HSS (purple),
for comparison.
The  fit results for the bursts in the LHS after the HSS (green) are also included.
}
\label{flux_tal}
\end{figure*}

By using the persistent flux at 1.5--30 keV and the burst duration $\tau$ as derived in Chenevez  et al. (2010), we  produced the $\tau$-flux diagram. This is shown in Fig. \ref{flux_tal}, where we also show linear fits (the parameter p1 enclosed in Fig. \ref{flux_tal} stands for the slope of these linear fits) to the data related to bursts occurring in different spectral states.
We find that the fits worsen when the bursts from HSS and number 8 in the classification of Chenevez  et al. (2010) are included. Recently,  Zhang  et al. (2010) found
while analyzing the cooling phase of type-I X-ray bursts in 4U 1636-53
that the bursts occurring in HSS may be He-dominated, while they could be H-dominated in LHS.
Burst number 8 occurred less than 10 minutes later after it predecessor, and  it  most likely has the same fuel (Fujimoto et al. 1987).
We therefore consider an overall sample for the bursts occurring in the LHS that excludes burst number 8. We proceeded to check the $\tau$-flux relationship in burst groups (the different color bars above the PCA lightcurves in the Fig. 1) located in the initial rising LHS, the LHS plateau preceding the HSS, the whole LHS preceding HSS (i.e., the sum of the first two groups), the LHS after the HSS, and the overall sample (see Fig. \ref{flux_tal}). The correlation factors are derived as 0.90, 0.32, 0.68, 0.75 and 0.35, respectively.
A linear fit results in a reduced $\chi^2$ $\sim$ 2.1 for the overall and  $\sim$ 1.2-1.8 for the rest. These values seem
acceptable because the overall sample consists of grouped bursts showing different slopes in this diagram, for which the  $\chi^2$ is individually better.
 We find that  the  bursts  in the decaying LHS present  a linear slope  of 5.0$\pm$1.0, which is  substantially different from  the 1.7$\pm$0.1 of the overall and  0.9-1.8 of the rest (see Fig. \ref{flux_tal}).
A parabolic line slightly improves the fit,  with  $\chi^2$ $\sim$62.61 (35 d.o.f.) for all the bursts except number 8. Fitting all bursts except number 8 with a broken line results in  a $\chi^2$ $\sim$53.08 (34 d.o.f.), and a probability of 0.00296 under an F-test for having a \textbf{marginally-significant} turn-over component at the lower flux.

\begin{figure*}
\centering
 \includegraphics[angle=0, scale=0.7]{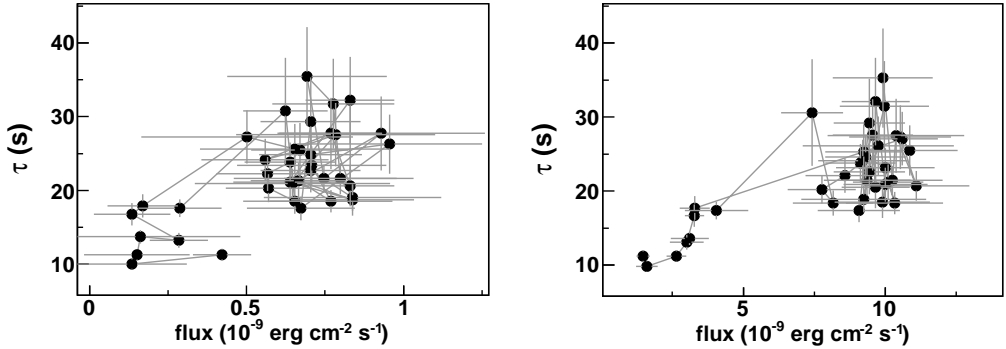}
      \caption{Plots of the timescale $\tau$ vs. the thermal (left panel with bolometric blackbody flux) and non-thermal (right panel with power-law component in the 1--100 keV band) persistent flux  for the bursts occurring in the LHS.
}
\label{pow_bb_flux_tal}
\end{figure*}

\begin{figure*}
\centering
   \includegraphics[angle=0, scale=0.4]{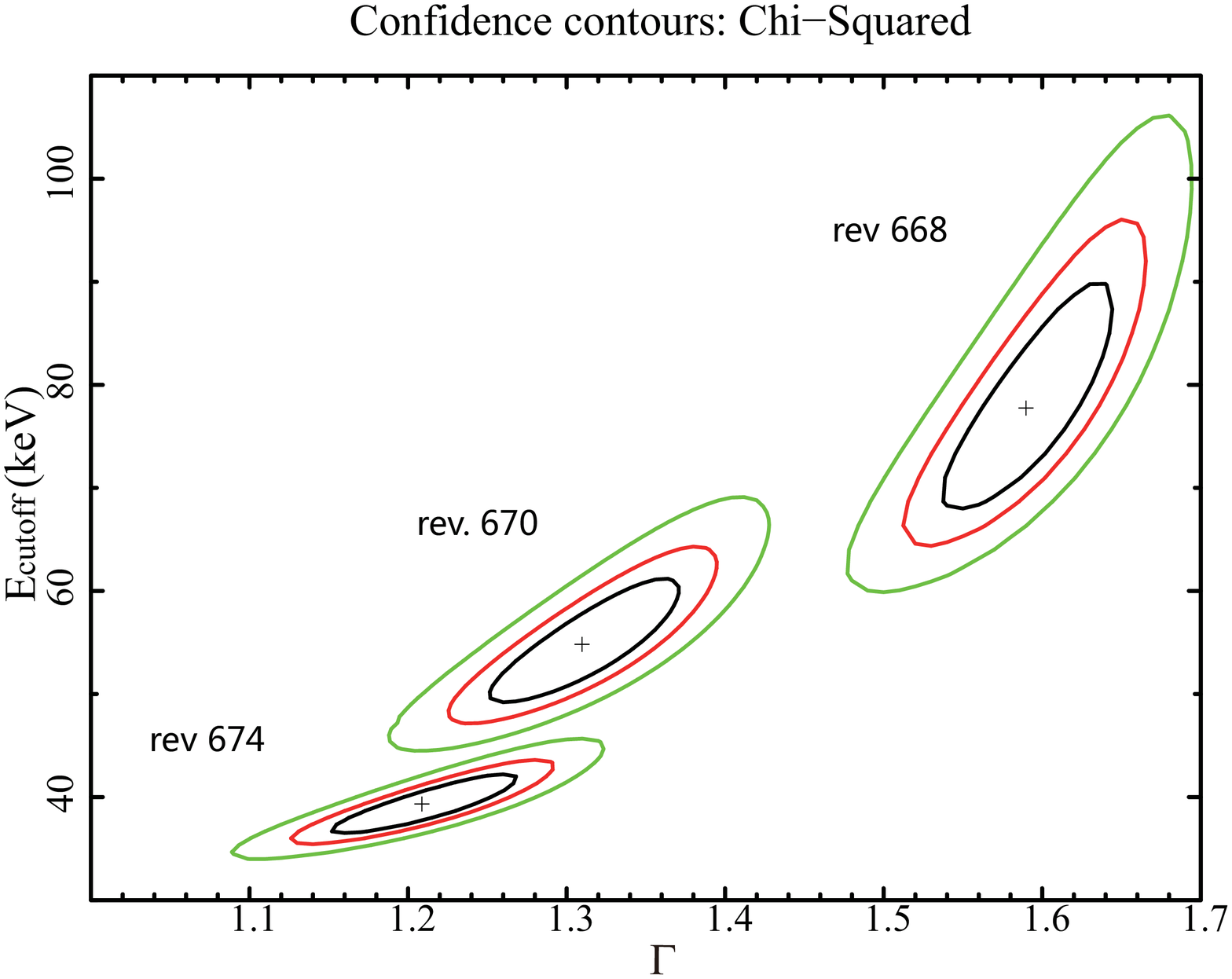}
      \caption{Six-eight percent (1$\sigma$, inner, black), 90\% (1.6$\sigma$, middle, red), and 99.7\%(3$\sigma$, outer, green) confidence level contours of the high-energy cut-off energy $E_{\rm cutoff}$ versus the photon index $\Gamma$ based on  the joint PCA/JEMX/HEXTE/ISGRI spectra of rev. 668, rev. 670 and rev. 674 (the last revolution of \emph{INTEGRAL} in the rising phase) of IGR~J17473--2721. The best-fit value is indicated by a ``+".}
\label{contour}
\end{figure*}

If we  enlarge  the energy band   to
0.1-200 keV, the $\tau$-flux correlation factors of the four burst groups described above  are 0.91, 0.29, 0.72, 0.35, and 0.52, respectively. A linear slope is derived as 0.7$\pm$0.2 for the leading LHS and 1.9$\pm$0.4 for  the decaying LHS. Thus, the trend still holds  that  the linear slope in the decaying LHS is steeper than in the leading LHS, although the persistent flux is estimated with an  energy band (0.1-200 keV) well beyond the PCA domain.
We also investigate the dependence of thermal and non-thermal emission, respectively, and
obtained correlation factors of 0.44 and 0.73 (see Fig. \ref{pow_bb_flux_tal}). Because the  thermal emission is
weak in the low/hard state, the error bars of the thermal flux turn out to be too large in Fig.
\ref{pow_bb_flux_tal}, which prevent us from  inferring the $\tau$-flux relationship further than a simple overall
positive correlation.   Moreover it is  hard to disentangle the disk emission  from the thermal emission from the
neutron star surface. But apart from this the diagram of  $\tau$ vs thermal flux as shown in Fig. \ref{pow_bb_flux_tal}
may be used as an standard candle in investigating the dependence of the burst duration on the accretion rate.

\section{Discussion}

\subsection{Spectral evolution from quiescent state to LHS}

The cutoff energy of the power-law component in X-ray spectra is a diagnostic parameter of the electron energy distribution in the
corona and of the analytical approximation of unsaturated Comptonization. The latter is
used extensively to study the X-ray evolution of BH and NS XRB
outbursts (e.g., Motta et al. 2009, Chen  et al. 2006).
In general,  the spectrum is typical of an LHS with a high energy cutoff at $\sim$ 40 keV for the NS binaries and $\sim$ 80 keV for the BH binaries.
 However, there are only a few sources observed in the X-ray band at the beginning of the outburst because either the transition is too quick or the source is too faint  in this period. This is especially true for the NS X-ray binaries because only the low-luminous-type atoll sources reach the LHS.
For instance, for the LHS of the outburst from Aql X-1 in 2005, which has an LHS before an HSS analogous to
IGR~J17473--2721, the cutoff energy is $\sim$ 30 keV and no cutoff was detected up to 150 keV in the following
HSS  (Rodriguez et al. 2006).

In the 2008 outburst of IGR~J17473--2721 we have found here that the cutoff energy $E_{\rm cut}$ gradually
decreases in the state transition from quiescence to LHS state within ten days, which is very similar to the spectral
evolution observed from BH X-ray binaries. Accompanied with the spectral evolution shown in our previous work
(Zhang et al. 2009), evidently sees that $E_{\rm cut}$ abruptly decreases from  $E_{\rm cut}=38.51^{+2.11}_{1.90}$ keV
to $\sim$5 keV when the source moves from the LHS plateau to HSS, within three days. For the LHS after the HSS, the
hard  tail of  spectra was fitted with a bknpow ($E_{\rm bkn}\sim$51 keV) or Comptt ($E_{\rm e}\sim$22 keV).
This complete evolution of the spectrum follows the sequence of  quiescent state $\rightarrow$ LHS $\rightarrow$
HSS $\rightarrow$ LHS in the outburst of IGR~J17473--2721.

From the observations of the transition from quiescent state $\rightarrow$ LHS in IGR~J17473--2721 with
\emph{RXTE} and \emph{INTEGRAL}, one sees that  the total flux is rising in the soft and hard X-ray bands (3--200
keV), the spectra become hard, and $E_{\rm cut}$ decreases monotonously. This resemblance in spectral behavior
between BH and NS binaries may be a clue for understanding the state transitions in both systems, which is related to
the cooling/heating of the corona by the photons produced at the accretion disk or/and the NS surface (Esin et
al. 1997; Menou et al. 1999).
However, after the initial LHS, spectral differences between BH and NS binaries occur. From our previous work
(Zhang et al. 2009), $E_{\rm cut}$ of IGR~J17473--2721 in HSS was measured to be $\sim$5 keV, similar to that of
4U~1608-522 ($E_{\rm cut}$ $\sim$3 keV in its HSS, as reported by Chen et al. 2006). These values are different
from BH systems (where no cutoff is observed in HSS). This inconsistency between the two systems may be related
to the existence of a hard surface or/and the magnetic field in NS binaries (Chen et al. 2006, Zhang et al
1996).

\subsection{Does most of the accretion material fall to the NS surface in the LHS?}\label{outflow}

The persistent emission is usually taken as an estimator of the accretion rate in XRB; but it is not always
easy to find out if a significant outflow reduces the final inflow rate onto the central compact object. It is
also not clear whether the inflowing material exhausts all its gravitational potential energy before reaching
the surface of an NS or the event horizon of a BH. Fortunately, under the assumption that the burst exhausts
the entire accreted material on the NS surface, bursts may be used to provide a good estimate for the actual inflowing
rate; this cannot be done easily for BH XRBs.

The accretion rate estimated from the observed persistent emission is given by $\dot M_{\rm p} \sim L_{\rm p}
/ e$, where $L_{\rm p}$ is the persistent luminosity and $ e$ is radiative efficiency. In NS binaries, the
potential energy released per nucleon (of mass $m_{\rm p}$) is $G M_{\rm NS}$$m_{\rm p}/R_{\rm NS}\approx200\
{\rm MeV}$ ($M_{\rm NS}=1.4\ M_{\odot}$, $R_{\rm NS}=10\ {\rm km}$), or $\sim 20\%$ of its rest mass. We then
have $\dot{M}_{\rm p} \sim {\rm 2.4}\times {\rm 10}^{\rm 17}$  g s$^{ -1}$ with $L_{\rm p} \sim$ 0.11 $L_{\rm
Edd}$ in the LHS preceding the HSS of IGR~J17473--2721.

On the other hand, the inflowing rate can be estimated from the bursts occurred in  the LHS preceding the HSS of  IGR~J17473--2721 by $\dot M_{\rm b} \sim L_{\rm b} / e$,
where $L_{\rm b} \sim  1.4 \times  10^{-3} \; L_{\rm Edd}$ is the average luminosity of the bursts in the LHS,
$e \sim Q_{\rm nuc}/938$=$({\rm 1.6}+{\rm 4}<X>)/$938, $Q_{\rm nuc}$ (in units of MeV) is the energy released per nucleon during the burst,
and $X$ is the hydrogen mass fraction of the fuel. $\dot M_{\rm b}\sim\dot M_{\rm p}$ requires $<X> \sim 0.18$, a
reasonable value and consistent with the durations of the observed bursts.
We therefore suggest that at least
for IGR~J17473--2721, most of the accreted material will fall to the NS surface in the LHS.

Fig. \ref{pow_bb_flux_tal} shows that, similarly to other NS binaries in the LHS, the power-law component
dominates the persistent flux in the two observed LHSs of IGR J17473--2721. This means that the majority of the
gravitational potential energy is released in the corona before reaching the surface of the NS; otherwise the
kinetic energy of the accreted material would produce luminous thermal (blackbody) radiation from the surface of
the NS.

\subsection{Spherical or disk corona ?}

The almost pure blackbody spectrum of the observed type-I bursts indicates   no  Comptonization of the
thermal photons from the surface of the neutron star, i.e., the corona does not cover a large fraction of the NS
surface.  The very different ratios between the observed thermal photons and the power-law photons in the
quiescent state and in the bursts suggest that the soft photons from the surface of NS do not dominate the cooling of
the corona.
 An additional evidence against spherical corona comes from the previous   spectral analysis of this outburst (Zhang et al. 2009), where the system was found to be highly inclined and most of the disk emission is not supposed to be visible.
Additional evidence for a disk corona is that a radius $\sim$3 km of the blackbody in the persistent flux (the fourth panel of Fig. \ref{spe})   is more compatible with being a hot spot on the NS surface than its inner accretion disk radius.

On the other hand, a corona on top of the disk has a relatively smaller coverage for the soft X-rays from the
surface of the NS, naturally avoiding all  the difficulties discussed above. In this case, the non-detection
or the low level of the disk's thermal X-ray emission would suggest that the corona covers the disk almost
completely and that a saturated inverse Comptonization process produces the observed power-law spectrum in the
persistent emission. We conclude that it is likely that most of the
thermal photons from the NS surface are not Comptonized, because the corona covers the disk, but not the
NS.

\subsection{Toward a ms pulsar system?}

By neglecting the influence of a possible outflow on the estimation of the accretion rate, assuming a radiative
efficiency of   $\sim$ 20\%  and a typical duty cycle of   $\sim $7\% (Sivakoff et al. 2008), the mass growing-rate of an accreting NS can be derived as
\begin{equation}
\delta M=\frac{L_{\rm bol}\tau  \delta}{\epsilon},
\end{equation}
where $L_{\rm bol}$ is average NS luminosity during outbursts, $\epsilon$ is the radiative efficiency, $\delta$
is the duty cycle, and $\tau$  is the lifetime of the NS binary. By taking $\tau\sim$${\rm 10}^{\rm 8}\ {\rm
yr}$ and $L_{\rm bol}\sim$  0.1--1$L_{\rm Edd}$, we find  that  the NS will accrete a mass of 0.02--0.22
$M_{\odot}$ from its companion. Millisecond pulsars (MSP) with  a spin  period ($P_{\rm s}<$20 ms) seem to have
a larger mass ($M_{\rm NS}$=1.57 $\pm$ 0.35 $M_{\odot}$) than the mean value of 61 measured masses of NSs (1.46
$\pm$ 0.30 $M_{\odot}$) (Zhang et al. 2011).
The estimation above may indicate that a X-ray burster such as  IGR J17473--2721  could
evolve toward an ms-pulsar system  (Demorest et al. 2010).

\subsection{Type-I X-ray bursts related to the spectral state of the outburst}

For the bursts embedded in the outburst, we found that for H-dominated burning the CNO cycle
constrains the speed at which the H fuel is exhausted, which may lead  to the burst duration  correlating with
the accretion rate.
The most recent analysis in Zhang et al. (2010) indicates that  the bursts at low hard state most likely have  H fuel. In  Fig. \ref{flux_tal} this correlation is present for the bursts in the overall LHS, and the LHS after HSS. The slope of these correlations  differs at a significance of $\sim$3 $\sigma$ level, with  the slope for the bursts occurring  in the decaying LHS being a factor of 3 steeper than that of the overall sample. An extrapolation of the slope for the bursts in the decaying LHS to the other three groups in Fig. \ref{flux_tal}  would overshoot the lower fluxes.
This feature may be related to part of the accretion material holding up to  form a corona on  its way of falling onto the surface of neutron star during the decaying part of the low hard state.
Therefore, the apparent  accretion rate on the surface of the neutron star prior to each burst would  be overpredicted if one simply were to take take the persistent emission as estimator.

\subsection{Summary}
With our complementary  analysis of \emph{INTEGRAL} and \emph{RXTE} observations of the 2008 outburst of IGR
J17473--2721 we find that

\begin{itemize}
\item The cutoff energy of the spectra decreases from $E_{\rm cut} \sim$ 150 keV to $\sim$ 40 keV  when the source
leaves the quiescent state toward the LHS, which is rarely reported in NS binaries. This phenomenon may be related to
the cooling/heating of the corona by the photons produced at the accretion disk or (and or) at the NS surface.

\item The corona seems to be located above the disk, but not around the NS, so that emission from  type-I
X-ray bursts can escape from the NS surface without suffering Comptonization.

\item  The different slopes of the linear relationship \textbf{hinted} in the burst duration  vs. the persistence flux diagram
for the LHS before and after the HSS  may be caused if part of the accretion material
is used to form a corona.

\end{itemize}

\acknowledgements
This work was subsidized by the National Natural Science Foundation of China, the CAS key Project KJCX2-YW-T03, 973 program 2009CB824800 and NSFC-11103020, 11133002. JMW and SZ thank the Natural Science Foundation of China for support via NSFC-10521001, 10733010, 10821061, 11073021, 11173023.
DFT acknowledges support from the grants AYA2009-07391 and SGR2009-811, as
well as the Formosa Program TW2010005.
This research has made use of data obtained from the High Energy Astrophysics Science Archive Research Center (HEASARC), provided by NASA's Goddard Space Flight Center.


\end{document}